\documentclass[aps,twocolumn,superscriptaddress]{revtex4}
\usepackage{amsfonts}
\usepackage{amsmath}
\usepackage{amssymb}
\usepackage{graphicx}
\usepackage{float}
\usepackage{color}
\begin{document}
\title{Simulation of a flat folding nano-swimmer confined in a nanopore. }

\author{Gaetan Delhaye}
\affiliation{Laboratoire de Photonique d'Angers EA 4464, Universit\' e d'Angers, Physics Department,  2 Bd Lavoisier, 49045 Angers, France}

\author{Felix Mercier}
\affiliation{Laboratoire de Photonique d'Angers EA 4464, Universit\' e d'Angers, Physics Department,  2 Bd Lavoisier, 49045 Angers, France}

\author{Victor Teboul}
\email{victor.teboul@univ-angers.fr}
\affiliation{Laboratoire de Photonique d'Angers EA 4464, Universit\' e d'Angers, Physics Department,  2 Bd Lavoisier, 49045 Angers, France}

\keywords{dynamic heterogeneity,glass-transition}
\pacs{64.70.pj, 61.20.Lc, 66.30.hh}

\begin{abstract}
We use molecular dynamics simulations to investigate the displacement of a simple butterfly-like molecular motor inside nanopores of various radii filled with a viscous medium.
The medium is modeled with a versatile potential that may be adjusted to represent  a large number of materials. 
It was found previously that the motor's folding not only increases its displacement but also creates elementary diffusion processes inside the medium, related to the opening angle of the motor 's folding. The presence of these processes changes the medium's dynamics and in turn affects the motor's displacement.
Therefore we test the motor's displacement with different activations of the medium inside the pore by varying the motor's opening angles. 
We find that the optima of the motor's displacement oscillate with pore sizes and that the optimal radii  depend  on the activation of the medium.
These results imply that it is possible to choose the activation or opening angle that optimizes the motor's displacement for a given pore size. 
Results also show that the activation decreases strongly the confinement's hindering of the motor's motion, in particular in small pores.
Finally, analyzing the distribution probability of the motor's position and the density of elementary motions we find that the motor is mainly located in the center of the pore.
We find  spikes in the density of elementary motions when the motor goes away from the center, suggesting important contributions of the motor's bouncing motions on the pore walls.

\end{abstract}

\maketitle
\section{ Introduction}

Molecular motors are ubiquitous in biological matter where they are responsible for the intracellular motions and catalysis that make life possible.
The idea of mimicking nature that lead to the rise of nanotechnology following the lecture of Richard Feynman\cite{feynman} while already prolific,  still faces the challenge to create synthetic nano-motors. 
For that reason the design and properties of nano-motors and nano-swimmers is an active field of research\cite{nanobook,mot20,moto1,moto2,moto3,moto4,moto5,moto6,moto7,motor0,motor1,motor2,motor3,motor4,motor5,motor6,motor7,motor8,motor10,motor11,motor12,motor13,motor14,motor15,motor16,motor17,pccp,motor18,nano1,dattler} with number of applications and theoretical perspectives expected. 
The applications range from medicine to engineering, sensing, the control of transport mechanisms, crystallization, hydrophobicity, catalyze and actuation to name a few.
 Important theoretical perspectives also appear due to the connection with out of equilibrium statistical physics and biophysics, leading to a fascinating new field of research called 
 active matter\cite{active1,active2,active3,active4} that develops rapidly.
 Synthetic molecular motors  are however still  relatively complex while the   
creation of simpler motors would be of large interest, permitting to produce them more easily and in larger numbers,  increasing moreover their stability. 
However the creation of nano-motors face a number of difficulties including Purcell's theorem\cite{scallop1,scallop2,prefold,scallop13b,scallop13c,scallop4,scallop7,scallop5,scallop6,scallop8,scallop9,scallop11,scallop12,scallop13,scallop14,scallop15,scallop10,scallop3,elec1} due to the two following  main physical reasons:
Brownian incessant bombardment of the motor hinders directionality in the motor's motion, while the huge increase of the relative importance of the viscosity at the nanoscale,  hinders the motor's motion itself.

In this work we use a flat motor that folds and unfolds periodically inside a nanopore. Experimentally our motor can be seen as a simplified representation of the azobenzene molecule or derivatives that do have the property of photo-isomerization and which still unsolved behavior in soft matter have been the subject of numerous researches\cite{azo1,azo2,azo2b,azo3,azo4,azo5,azo6,azo7,azo8,azo9,flu5,cage,md16}. 
When confined inside a nanopore we expect the motor to be guided, leading to an increase of its orientation and displacement for long time scales.
Confinement has also been shown to induce a number of strange effects in supercooled liquids\cite{conf1,conf2,conf3,conf4,conf5,conf6,conf7,conf8,conf9,conf10,conf11,conf12,conf13,conf14,conf15,conf16} as a structural layering in the local density and large modifications of the dynamical properties of the liquid, mostly a slowing down and an increase of the viscosity.  
Confinement of active matter  offers  numerous  applications in particular for catalysis processes. 
As in biological matter, motor proteins are mostly moving in confined environments, the study of the interplay between confinement and active matter may also shed some light on biological motors behavior.

In a previous work \cite{aa} we have shown that our motor's opening angle was representative of the activation induced in the pore, and 
we will use that result in this work to test the effect of different activations of the medium on the motor's displacements.
Does the motor moves faster inside the pore than in the bulk ?
Does the motor moves faster when the medium is activated ?  
Interestingly enough, we find that the optima of the motor's displacements oscillate with pore sizes and depend on the activation of the medium.
We interpret these oscillations as a consequence of  different layering of the medium inside different pores and for different activations, however bouncing of the motor on the walls of the pore may contribute.
The activation decreases strongly the confinement's hindering of the motor's motion, in particular in small pores.
We also find  bouncing motions on the pore walls with important contributions on the motor's displacement, a result reminiscent of much more complex biological motor proteins.

\section{Calculation}

Together with theoretical and experimental methods, molecular dynamics and Monte Carlo simulations\cite{md1,md2,md2b,md4} and model systems\cite{ms1,ms2,ms3,ms4,ms5}  are now widely used to unravel unsolved problems in condensed matter and complex systems physics\cite{keys,md3,md4b,md6,md7,md8,md9,md10,md11,md12,md13,md14,md15,finite1,u1}. 

Our medium is modeled with a versatile potential\cite{ariane} created by joining together the two sort of atoms of the extensively studied Kob Andersen potential\cite{kob,kob1,kob2} and then optimizing the relative Lennard-Jones (LJ) parameters to still hinder crystallization. Due to its versatility, and the use of LJ potentials for the motor also,  our simulations can be translated to model although approximately a large number of materials.
For clarity we remind the versatility parameters at the end of that section.

\begin{figure}
\centering
\includegraphics[height=12. cm]{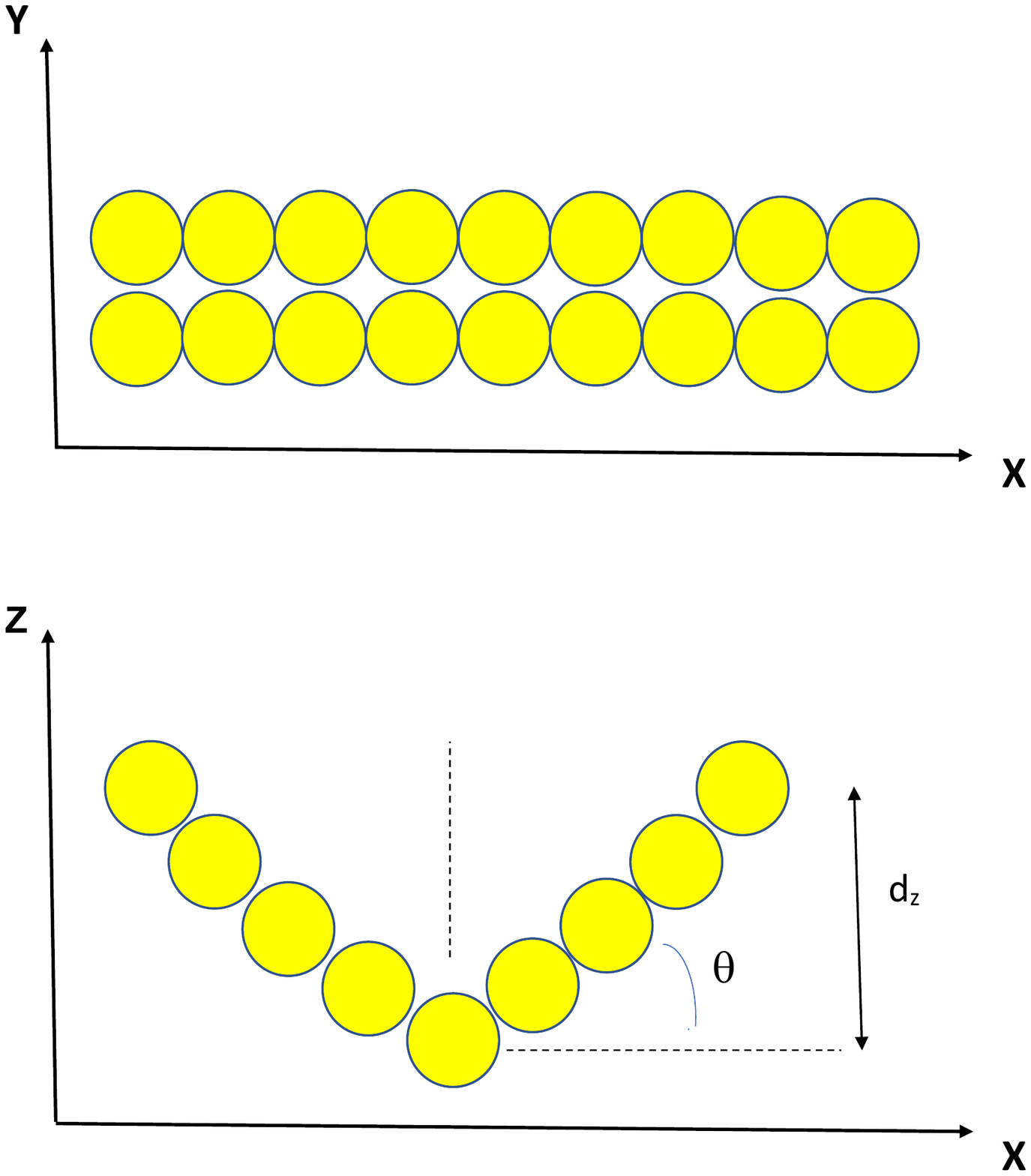}
\caption{(color online) Picture of the folded and unfolded motor molecule. Through this work we use the parameter $d_{z}$  to tune and quantify the activation of the medium by the motor's folding. That choice was validated in a previous paper \cite{aa}.
We call $X$,$Y$,$Z$  the laboratory coordinates that coincide with the local coordinates at origin time $t_{0}$ and are displayed on the Figure at two different origin times $t_{0}$.
We also define pore coordinates $x$,$y$,$z$ where $z$ is the pore axis.}
\label{f0}
\end{figure}

In our simulations we use  one motor molecule  (see Figure \ref{f0} for its description) imbedded inside a medium constituted of $1760$ linear molecules, in a parallelepiped box  $31.1$ \AA\ wide and $62.2$ \AA\ high. 
After aging the simulation box during $10 ns$ at the temperature of studies, we create the pore by suddenly freezing molecules located at a distance larger than $R$ from the center axis of our simulation box.
This procedure is intended to minimize the structure modification of the liquid by the confinement.
After their sudden freezing, the glassy molecules of the pores are no more allowed to move, leading to pure elastic interactions with the confined liquid.
Consequently the liquid temperature is not modified by the interactions with the pore walls.
After that procedure, as the dynamics changes with the confinement, the liquid is again aged during $10 ns$ before the beginning of any study.
We integrate the equations of motion using the Gear algorithm with a quaternion decomposition\cite{md1}  and a time step $\Delta t=10^{-15} s$. 
When the motor is active our simulations are out of equilibrium, because the motor's folding periodically release energy into the system.
However our system is in a steady state and does not age as explained below.
For that purpose we use the Berendsen thermostat\cite{berendsen} to evacuate from the system the energy created by the motor's folding. 
Notice that the thermostat applies to the medium's molecules but not to the motor. Thus the motor's cooling results only from the molecular interactions with the medium's molecules.
We use the NVT canonic thermodynamic ensemble as approximated by Berendsen  thermostat   (see ref.\cite{finite2} for an evaluation of the effect of the thermostat on our calculations), and periodic boundary conditions. 
 The molecules of the medium (host)\cite{ariane} are constituted of two rigidly bonded atoms ($i=1, 2$) at the fixed interatomic distance  {\color{black} $l_{h}$}$=1.73 $\AA$ $. These atoms interact with atoms of other molecules using the following Lennard-Jones potentials:
 \begin{equation}
V_{ij}=4\epsilon_{ij}((\sigma_{ij}/r)^{12} -(\sigma_{ij}/r)^{6})   \label{e1}
\end{equation}
with the parameters\cite{ariane}: $\epsilon_{11}= \epsilon_{12}=0.5 KJ/mol$, $\epsilon_{22}= 0.4 KJ/mol$,  $\sigma_{11}= \sigma_{12}=3.45$\AA, $\sigma_{22}=3.28$\AA.
The mass of the motor is $M=540 g/mole$ (constituted of $18$ atoms, each one of mass $30g/mole$) and the mass of the host molecule is $m=80 g/mole$ ($2$ atoms with a mass of $40g/mole$ each).
We model the motor with $18$ atoms in a rectangular shape constituted of two rows of  $9$ rigidly bonded atoms. 
The width of the swimmer is $L_{s}=4.4$\AA\ and its length $l_{s}=15.4$\AA. The length of the host molecule is $l_{h}=5.09$\AA\ and its width $L_{h}=3.37$\AA.
The motor's atoms interact with the medium's atoms using mixing rules and a Lennard-Jones interatomic potential on each atom of the motor, defined by the parameters: 
$\epsilon_{33}= 1.88 KJ/mol$,  $\sigma_{33}=3.405$\AA. 
We use the following mixing rules \cite{mix1,mix2}: 
\begin{equation}
\epsilon_{ij}=(\epsilon_{ii} . \epsilon_{jj})^{0.5}  ; 									
\sigma_{ij}=(\sigma_{ii} . \sigma_{jj})^{0.5}   \label{e2}
\end{equation}
  for the interactions between the motor and the host atoms. 
Our medium is a fragile liquid\cite{fragile1,fragile2}  that falls out of equilibrium in bulk simulations below $T=340 K$, i.e. $T=340$ K is the smallest temperature for which we can equilibrate the bulk liquid when the motor is not active. 
Consequently  the bulk medium above that temperature behaves as a viscous supercooled liquid in our simulations while below it behaves as a solid (as $t_{simulation}<\tau_{\alpha}$). 
Notice however that when active, the motor's motions induce a fluidization of the medium around it\cite{md16,flu1,flu2,flu3,flu4,carry,flu5,rate}, an effect that has been experimentally demonstrated\cite{flu1,flu2,flu3,flu4,flu5} with  azobenzene  photo-isomerizing molecules embedded in soft matter, a result that has been linked\cite{dh1,dh2} to the cooperative dynamics\cite{dh0} arising at the approach of the glass transition. 
We evaluate the glass transition temperature $T_{g}$ in the bulk to be slightly smaller  $T_{g} \approx 250 K$.
Notice that as they are modeled with Lennard-Jones atoms, the host and motor potentials are quite versatile.
Due to that property, a shift in the parameters $\epsilon$ will shift all the temperatures by the same amount, including the glass-transition temperature and the melting temperature of the material.
Each folding is modeled as continuous, using a constant quaternion variation, with a folding time $\tau_{f}=0.3 ps$.
The total cycle period is constant and fixed  at $\tau_{p}=400 ps$.
The folded and unfolded part of the cycle have the same duration equal to $\tau_{p}/2=200 ps$ including the $0.3 ps$ folding or unfolding.
Our system, while out of equilibrium, is in a steady state and is not aging. That behavior is obtained because the energy released by the motor into the medium is small enough and the time lapse between two stimuli large enough for the system to relax before a new stimuli appears.
In other words we are in the linear response regime\cite{pccp}.
Through this work we use the mean square displacements of the motor to measure its ability to diffuse.
The mean square displacement is defined as\cite{md1}:\\

\begin{equation}
\displaystyle{<r^{2}(t)>= {1\over N.N_{t_{0}}} \sum_{i,t_{0}} \mid{{\bf r}_{i}(t+t_{0})-{\bf r}_{i}(t_{0})} }\mid^{2}   \label{e130}   
\end{equation}

Notice that  as there is only one motor in our simulation the summation on $i$ can be eliminated for the motor. In that case, the mean square displacement is only averaged on the time origins $t_{0}$.

From the time evolution of the mean square displacement we then calculate the diffusion coefficient $D$  for diffusive displacements using the Stokes-Einstein equation:
\begin{equation}
\displaystyle{\lim_{t \to \infty}<r^{2}(t)>=2 p D t}
\end{equation}
where $p$ is the dimensionality. At short and intermediate timescales $p=3$, but as the motor will move at long time scales mostly in the direction of the pore axis, $p$ will tend to $1$.
To avoid this problem, we calculate the diffusion coefficient from the displacement in the direction of the pore axis $z$ only:
\begin{equation}
\displaystyle{\lim_{t \to \infty}<z^{2}(t)>=2  D t}
\end{equation}

\begin{figure}
\centering
\includegraphics[height=5. cm]{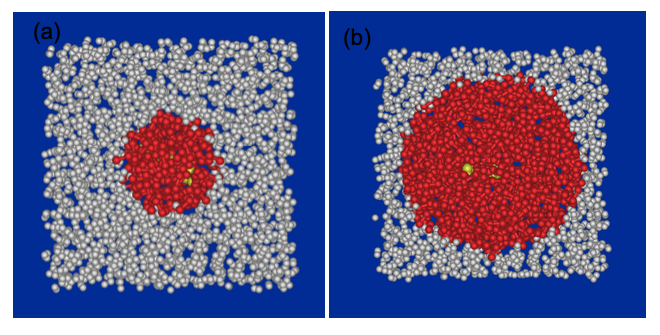}
\caption{(color online) Snapshots  in the $xy$ plan, of simulations for pores of radii (a) $R=6$\AA, and (b) $R=12.5$\AA. The colors are arbitrary.  
Our simulations use $8$ pores of radii varying from $5$ \AA\ to $12.5$ \AA\ separated by steps of $1$ \AA\ and $1.5$ \AA\ for the last step.
}
\label{f01}
\end{figure}

\begin{figure}
\centering
\includegraphics[height=8. cm]{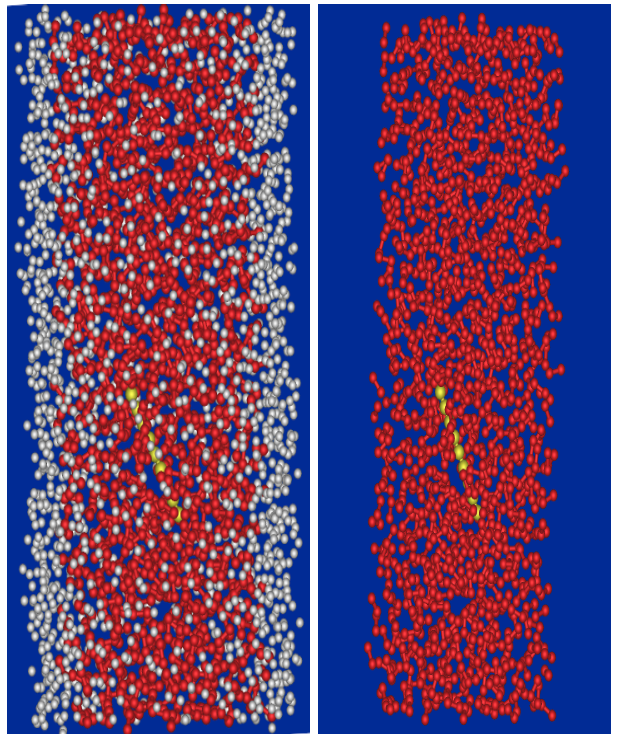}
\caption{(color online) Snapshot of a simulation observed from the side ($xz$ plan) to show the motor. On the right we have removed the pore walls from the picture. The pore radius is $R=12.5$\AA. The atoms sizes are arbitrary.}
\label{f02}
\end{figure}

Unless otherwise specified the results presented in this work correspond to a temperature $T=500 K$.
Notice however that our system is a model system with Lennard-Jones interactions only.
Therefore, due to the properties of the Lennard-Jones potential, one can shift the temperatures by any factor $\alpha$ ($T'=\alpha T$) creating a new material.
That new material will have all the $\epsilon$ parameters of the LJ potentials (including for the motor) modified in the same way $\epsilon ' = \alpha \epsilon$.
The results corresponding to that new material will have to be translated from the results of our work by a shift of time $t'=t / \alpha^{0.5}$.
Similarly, the size of the atoms can be scaled by a factor $\gamma$ ($\sigma'=\gamma \sigma$) provided that all the distances of the system are scaled by the same factor.
Our results will in that case have to be translated by a shift of distance $\gamma$ ($r'=\gamma r$) and a shift of time $\gamma$ ($t'=\gamma t$).
In that way our simulations are valid for a large number of materials although approximately, and experimentalists can adjust our data to their system of concern.

\section{Results and discussion}

\subsection{Dependence of the motor's displacement with pore radius}

\begin{figure}
\centering
\includegraphics[height=6 cm]{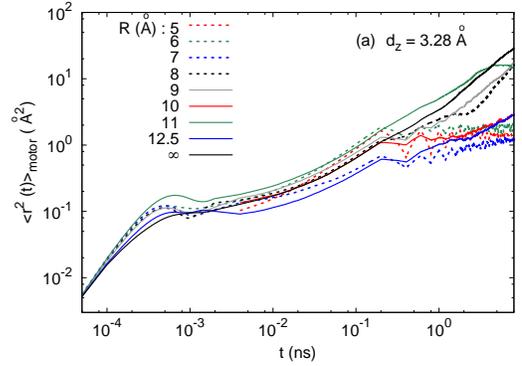}
\includegraphics[height=6 cm]{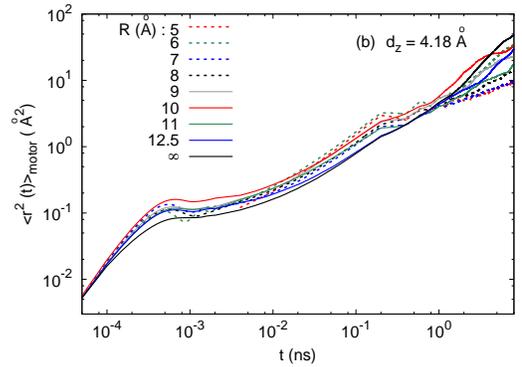}
\includegraphics[height=6 cm]{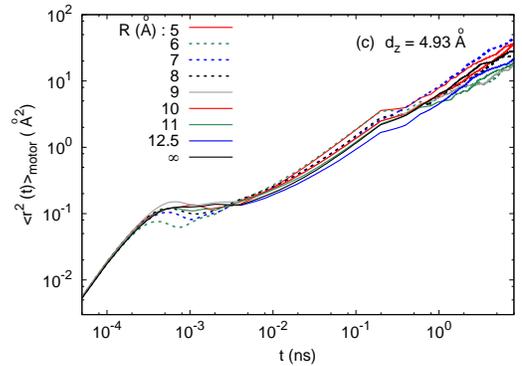}

\caption{(color online)  Motor's mean square displacements for different pore radii $R$.
The mean square displacement plotted here are calculated in the direction of the pore axis, in order to avoid any long term modification of $<r^{2}(t)>$ induced by restricted geometry.
 It is then multiplied by a factor $3$ to be comparable with 3-dimension mean square displacements in the bulk.
Each Figure corresponds to a different activation coefficient $d_{Z}$.
}
\label{ff6}
\end{figure}

\begin{figure}
\centering
\includegraphics[height=6.5 cm]{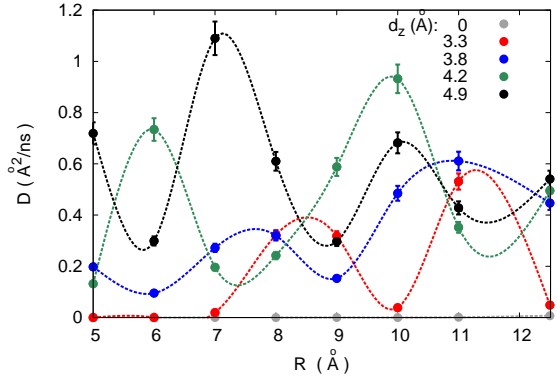}
\caption{(color online)   Diffusion coefficient of the motor as a function of the pore radius and activation parameter $d_{Z}$.
Note that the diffusion coefficients are displayed in a linear scale to be able to observe relatively small differences.
We observe oscillations in the diffusion coefficient, showing that some radii promote motor's diffusion.
}
\label{ff1}
\end{figure}

Figure \ref{ff6} shows the mean square displacements of the motor for different pore radii $R$, and activation $d_{Z}$. 
We observe that the activation mostly compensates the slowing down induced by confinement inside the pores.
More precisely, for the smaller activation $d_{Z}=3.28$ \AA, we observe two sorts of behavior.  For some pore radii, the motor diffuses very slightly, while for others it diffuses.
We note, that if the pore radii where the motor doesn't move are mostly the smallest, it is unexpectedly not always the case and some large pores have their motions hindered too. 
For larger activations $d_{Z}=4.18$ and $4.95$ \AA, the first behavior (motor hindered) disappears progressively as the activation increases, leading to a diffusion that appears to depend only weakly on the pore radius. However smaller variations may be hidden due to the logarithmic scale of the Figure.
Therefore, to study the evolution of the diffusion more accurately, we display in Figure \ref{ff1} the diffusion coefficient of the motor versus pore radius, for the different activation studied.
The diffusion coefficient is displayed in a linear scale to observe effects that cannot easily be seen on the previous Figures.

We observe oscillations in the diffusion coefficient, showing that some radii promote motor's diffusion while others hinder it.
These oscillations strongly depend on the activation $d_{Z}$. 
 In a previous work we found that the layering depends on the activation, an effect that probably originates from a modification of the contact properties with the pore walls for highly mobile molecules.
Therefore, the activation dependence of the oscillations of motor's displacement with pore radii, suggests that the oscillations originate from the medium's layering.

The distance between two peaks is approximately constant around $3$ \AA\  whatever the activation. 
Notice that the distance of $3$ \AA\  corresponds approximately to twice the distance between layers of the medium, a result that we understand as due to the large size of the motor.
 Actually, due to its size the motor is mostly embedded in two different medium's layers.  
These results therefore come in support of the picture that the oscillations of the motor's displacement are induced by the different layering oscillations of the different pores.

\begin{figure}
\centering
\includegraphics[height=6.5 cm]{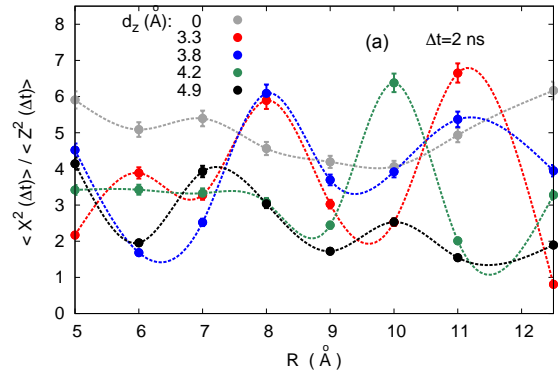}
\includegraphics[height=6.5 cm]{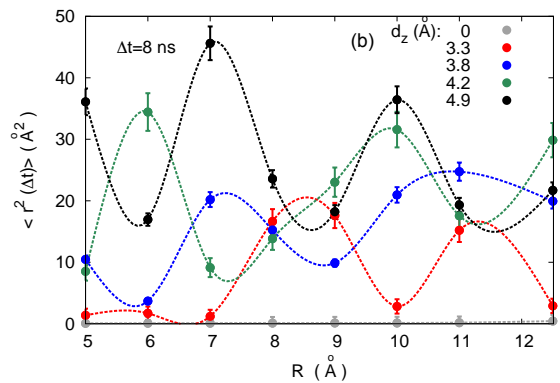}
\caption{(color online)  (a) Ratio of the mean square displacements along and perpendicular to the motor's axis versus the pore radius $R$ for a time lapse $\Delta t=2 ns$.
The orientation of the motor's displacements displays oscillations as a function of $R$.
(b) The mean square displacements oscillations for a time lapse $\Delta t=8 ns$ confirm the oscillatory behavior observed for the diffusion coefficient. 
}
\label{ff2}
\end{figure}

Because the diffusion coefficient calculation involves the evaluation of the slope of the mean square displacement, an evaluation that may induce uncertainties and some subjectivity, 
we verify these results in Figure \ref{ff2} with the motor's mean square displacement for two different time lapses $\Delta t= 2 ns$ and $8 ns$.
The Figure shows that some peaks of displacements are already present at short time scales, while others only develop at larger time scales. 
Eventually we observe at long time scale the same behavior than previously seen for the diffusion coefficient, confirming the preceding result.

\subsection{Dependence of the motor's displacement with activation}

\begin{figure}
\centering
\includegraphics[height=5.5 cm]{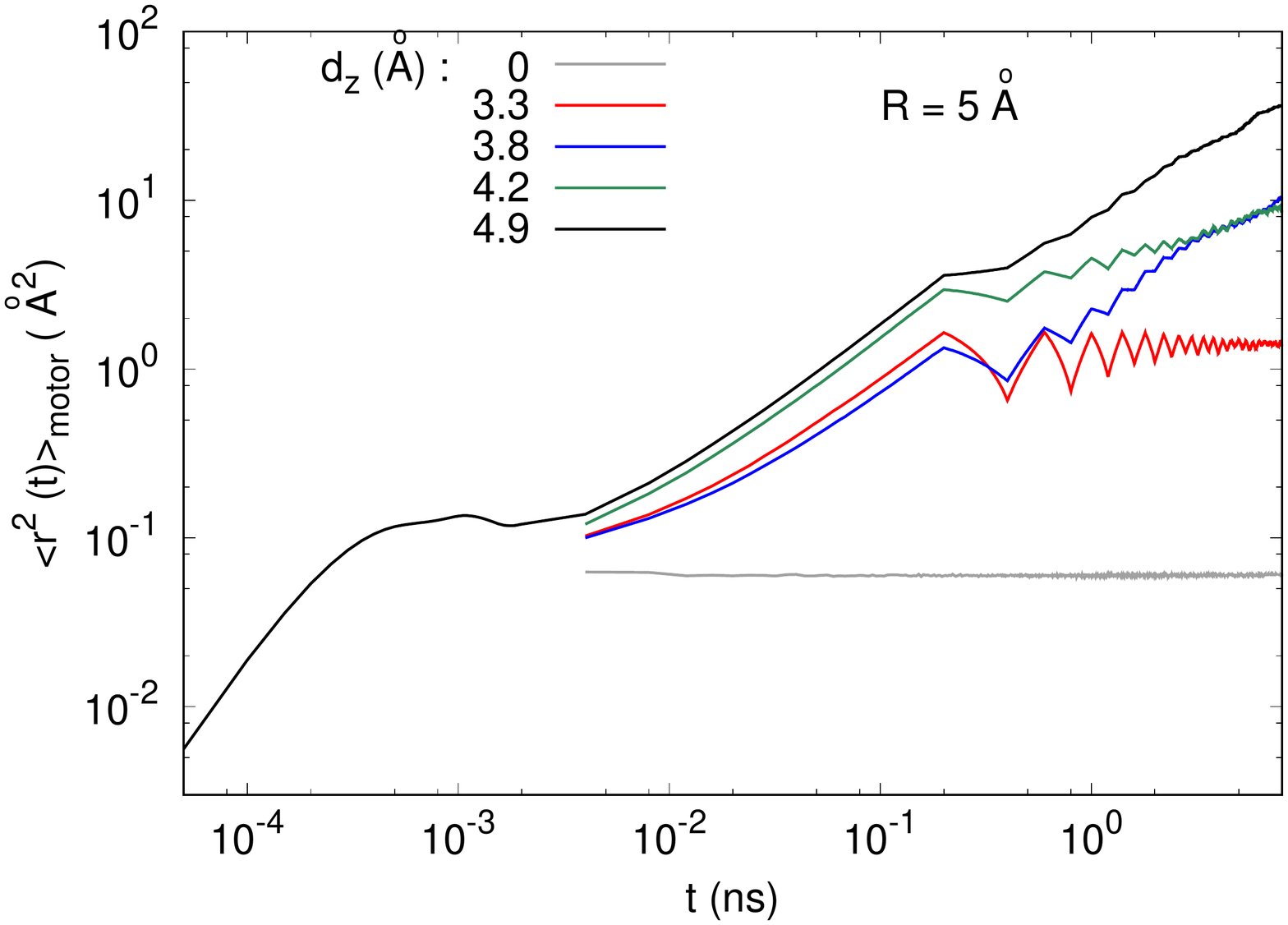}
\includegraphics[height=5.5 cm]{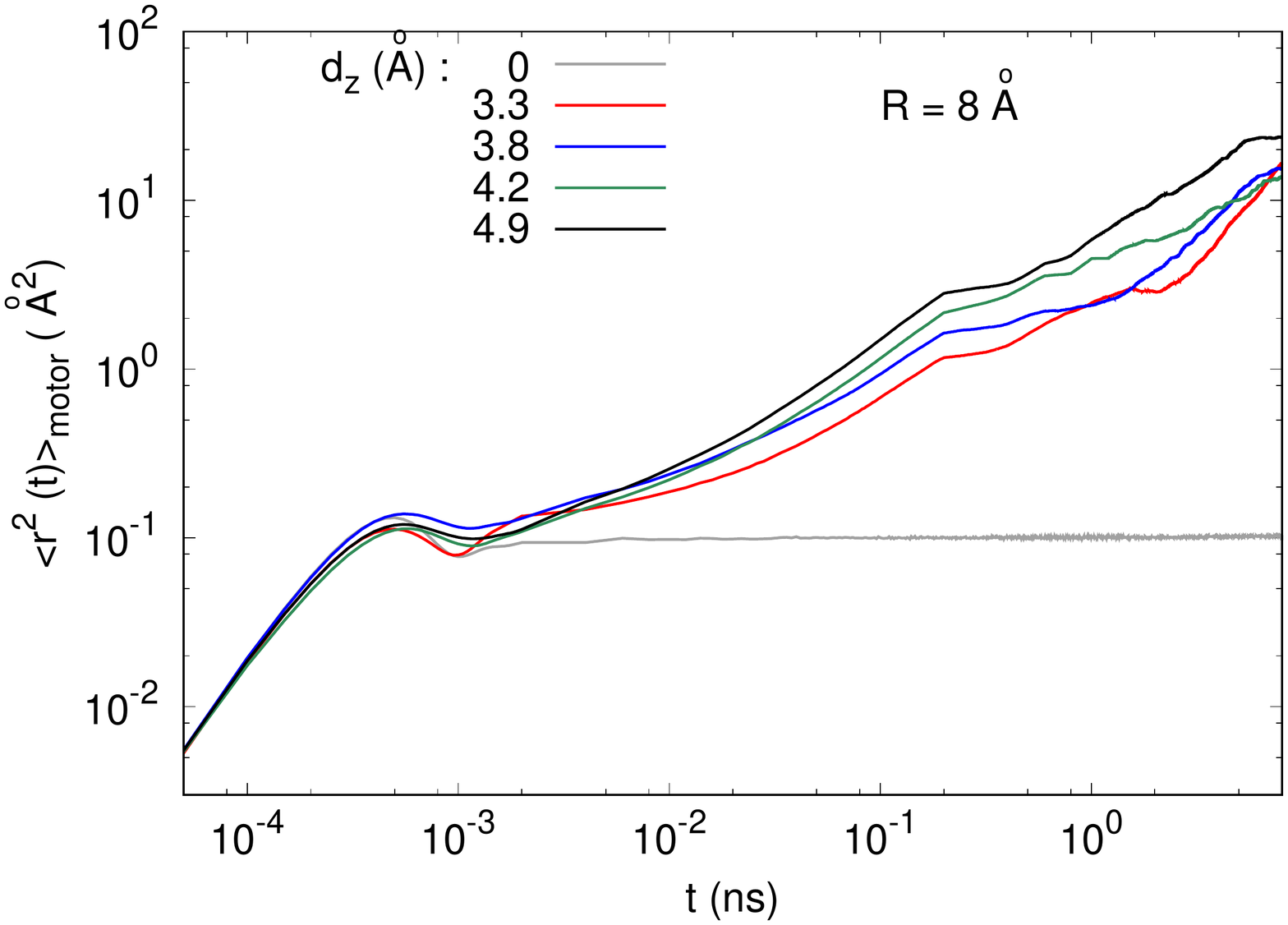}
\includegraphics[height=5.5 cm]{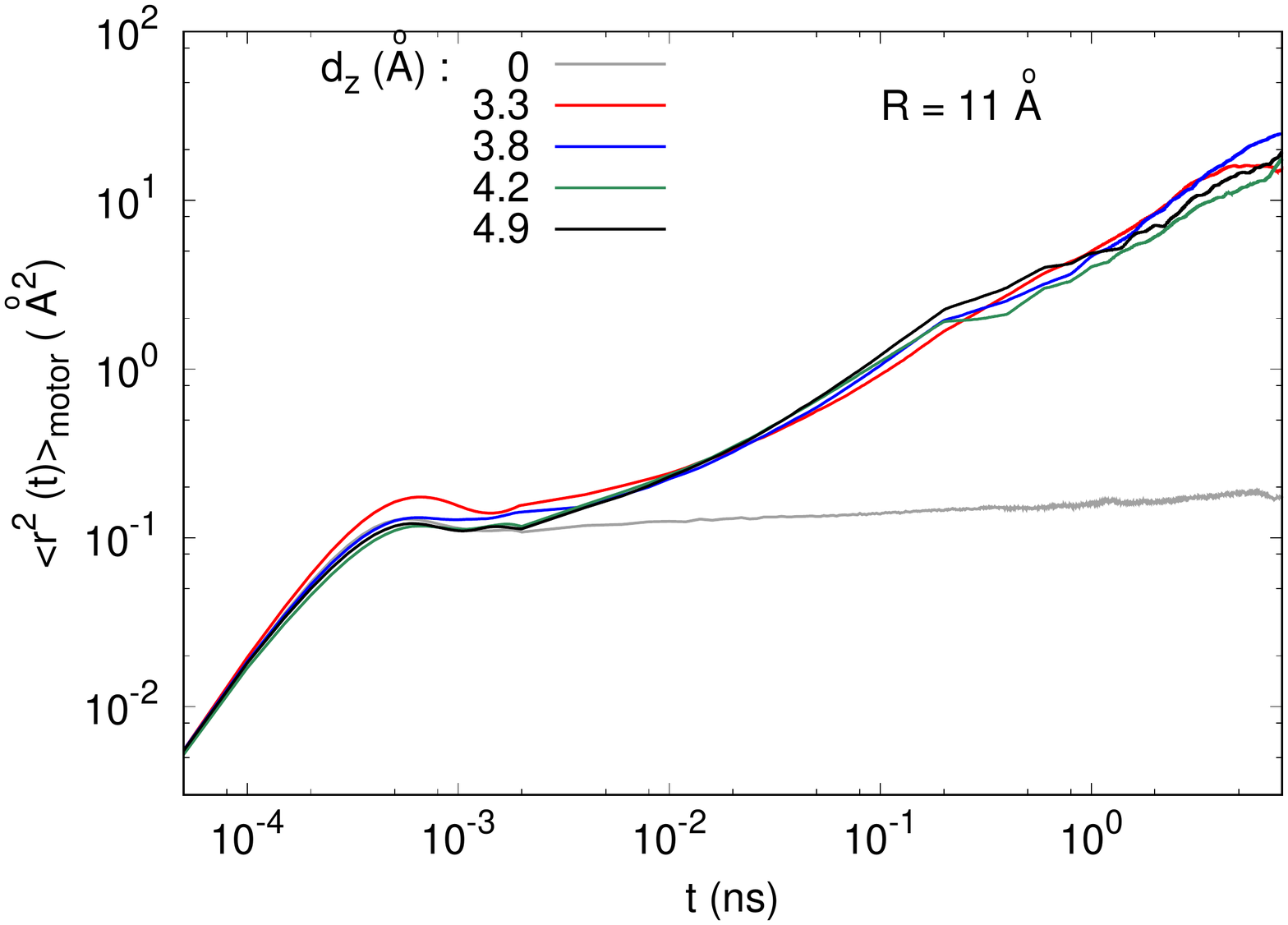}
\includegraphics[height=5.5 cm]{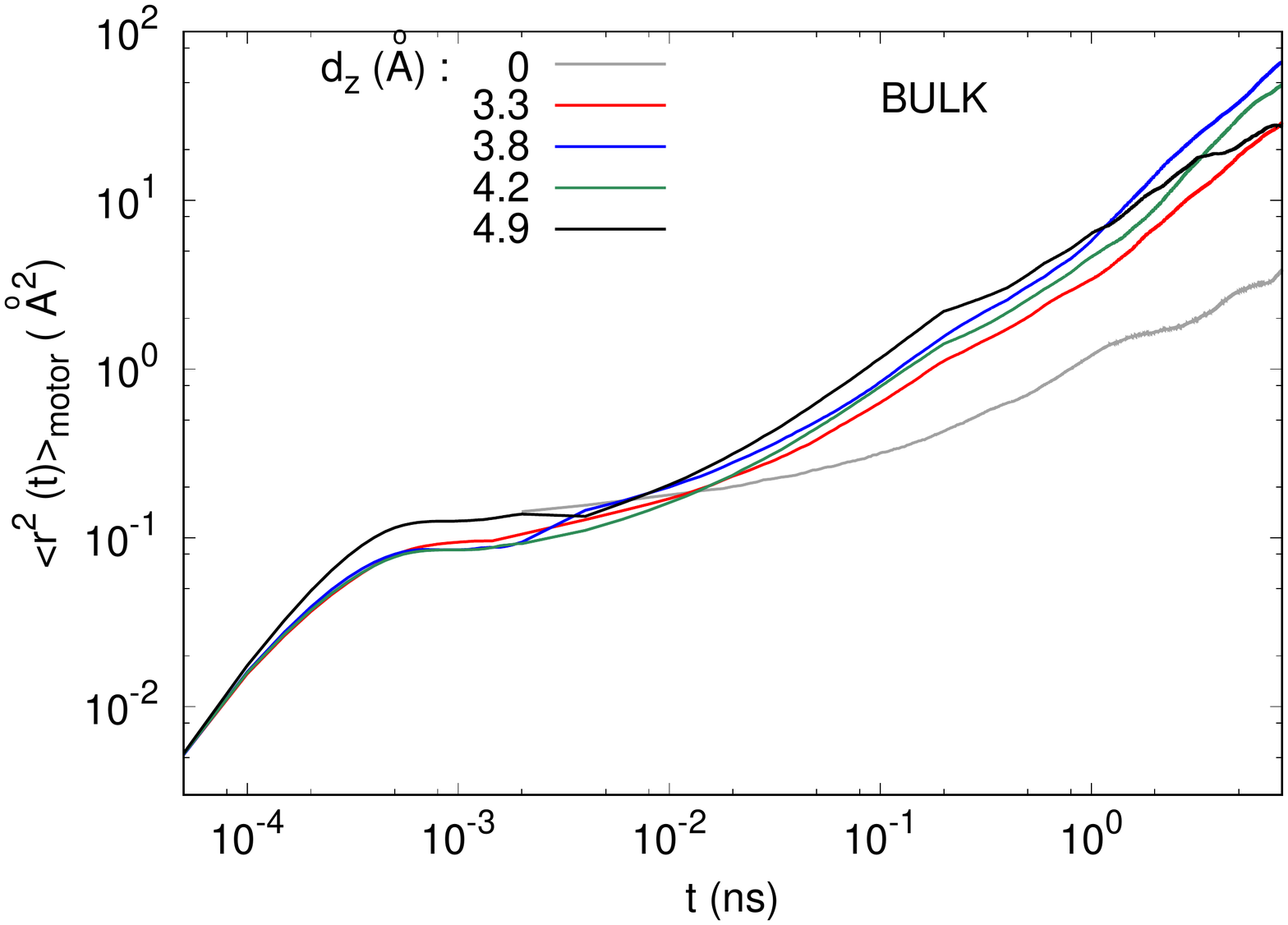}
\caption{(color online)  Motor's mean square displacement in the direction of the pore axis x$3$ for constant radii and different activations.
}
\label{ff7a}
\end{figure}

We now turn our attention to the effect of the activation on the motor's displacement.
For that purpose, Figure \ref{ff7a} displays the effect of the activation on motor's displacement for different pore radii.
For a pore radius $R=5$ \AA\ we observe a large dependence of the displacement with the activation,  the displacement increasing with activation.
Then as the pore becomes larger, that dependence decreases, while the gap in displacement between no activation and our smallest activation persists.  
For large pores a small activation is thus enough while larger activations do not improve significantly the motor's displacement. 
In contrast, for small pores a large activation results in a motor's displacement that appears no more hindered by the confinement slowing down.

Finally, we note that the activation modifies the oscillations of the displacements versus pore radius in Figures \ref{ff1} and \ref{ff2}, and consequently the optimal radii for displacement.
For a given porous material, the activation can therefore be chosen to optimize the motor displacement.

\subsection{Orientation of the motor's displacement within the pore}

A possible origin of the oscillations of displacement comes from the orientation of the motor inside the pore.
In the bulk, at short time scales anisotropic molecules move preferentially in the direction corresponding to their smaller surface.
That preferential direction of motion is then washed out by Brownian motion at larger time scales.
However due to their ability to move at low temperature where Brownian motion is weak, anisotropic molecular motors can have a preferential direction of motion for longer times than passive molecules.
In our case, as discussed in a previous paper \cite{previous}, the preferential direction corresponds to the $X$ axis (see Figure \ref{f0}).

\begin{figure}
\centering
\includegraphics[height=5.5 cm]{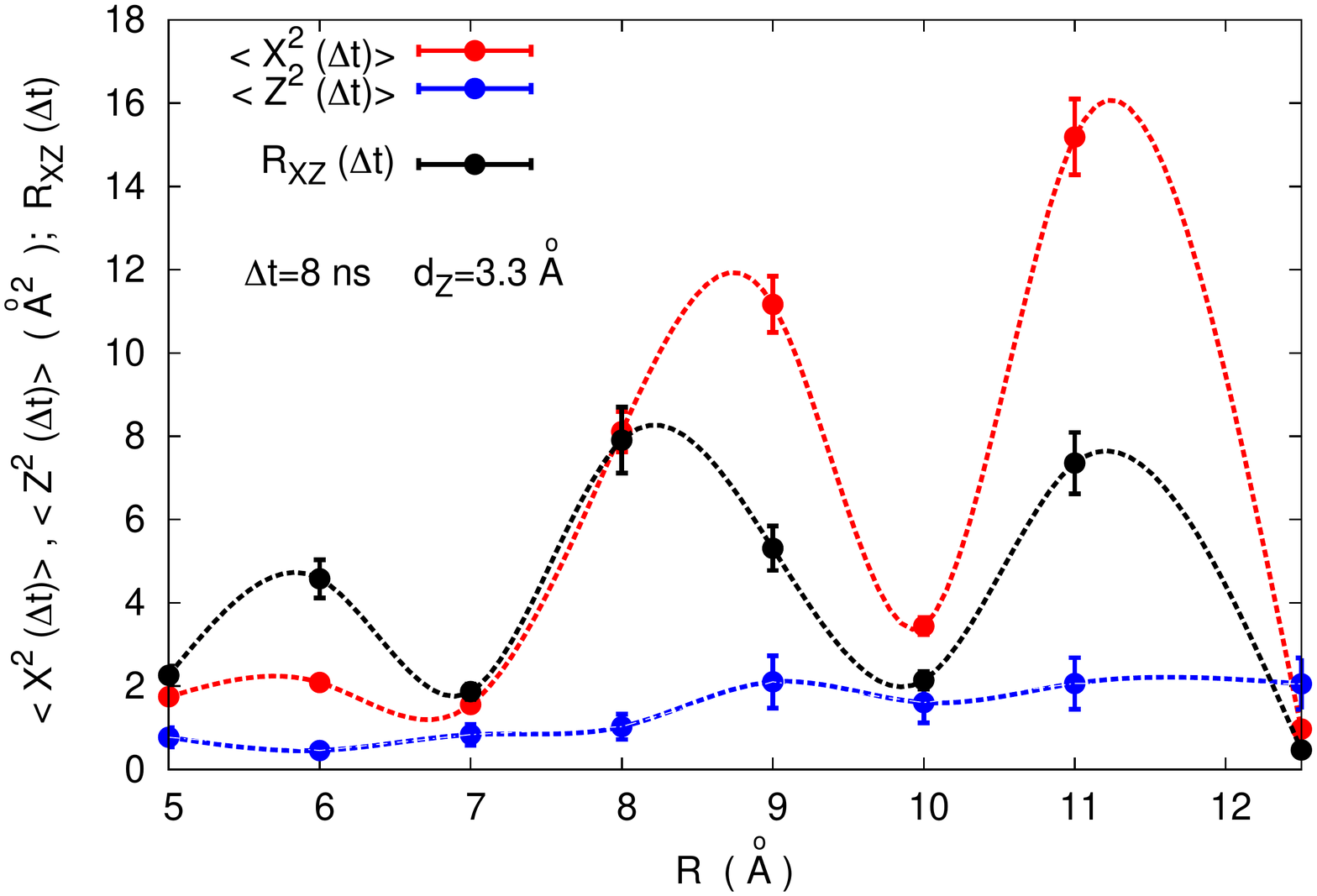}
\includegraphics[height=5.5 cm]{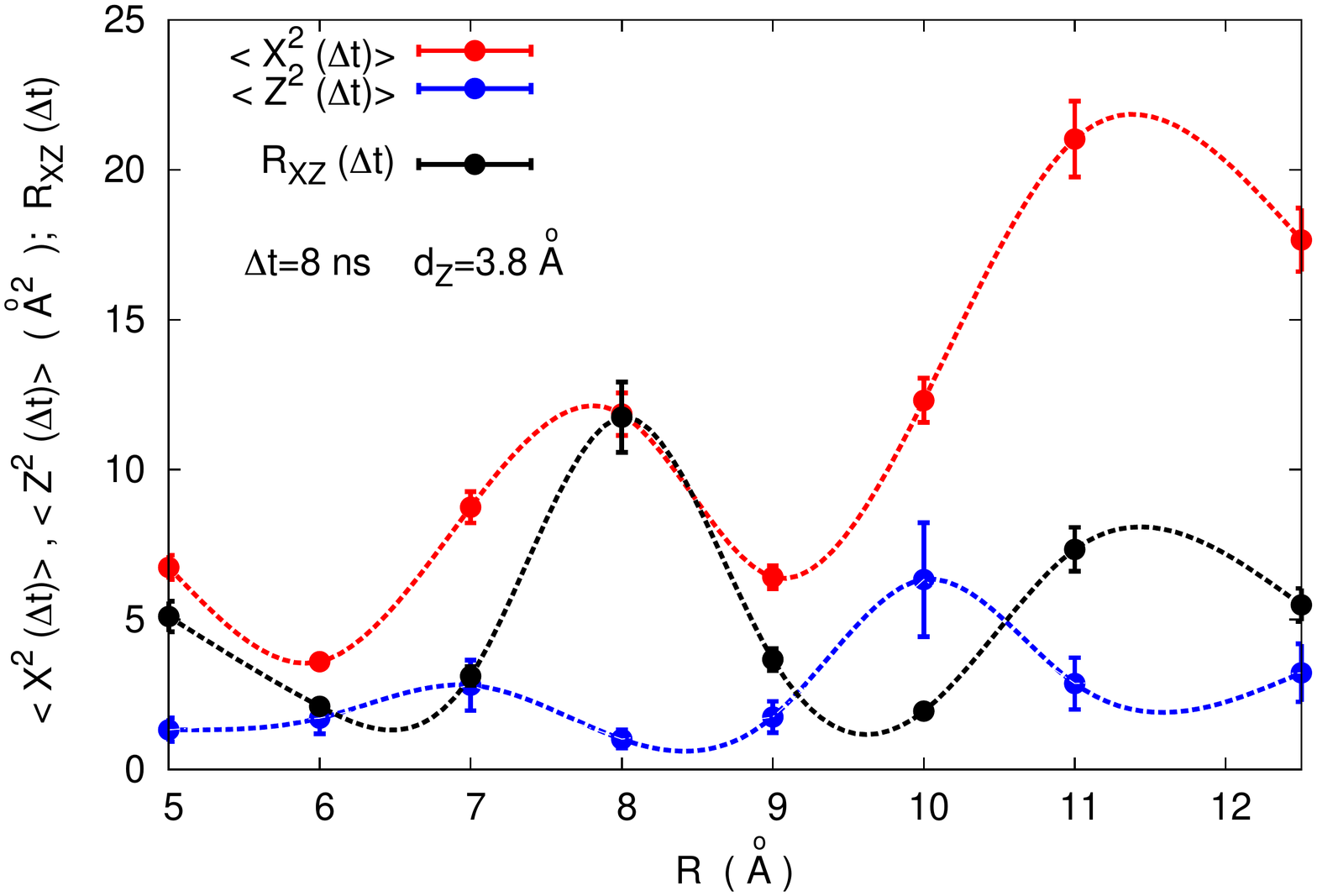}
\includegraphics[height=5.5 cm]{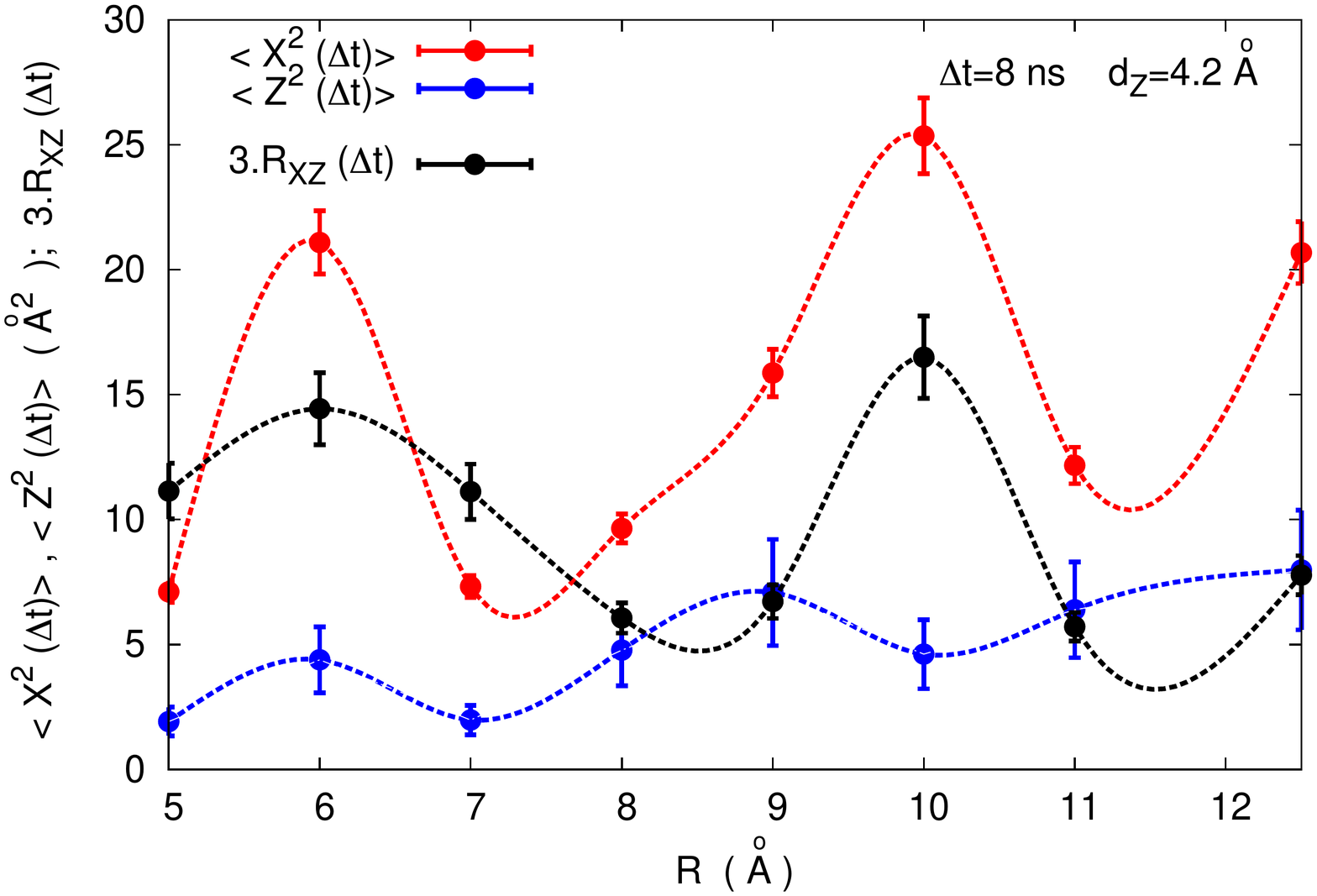}
\includegraphics[height=5.5 cm]{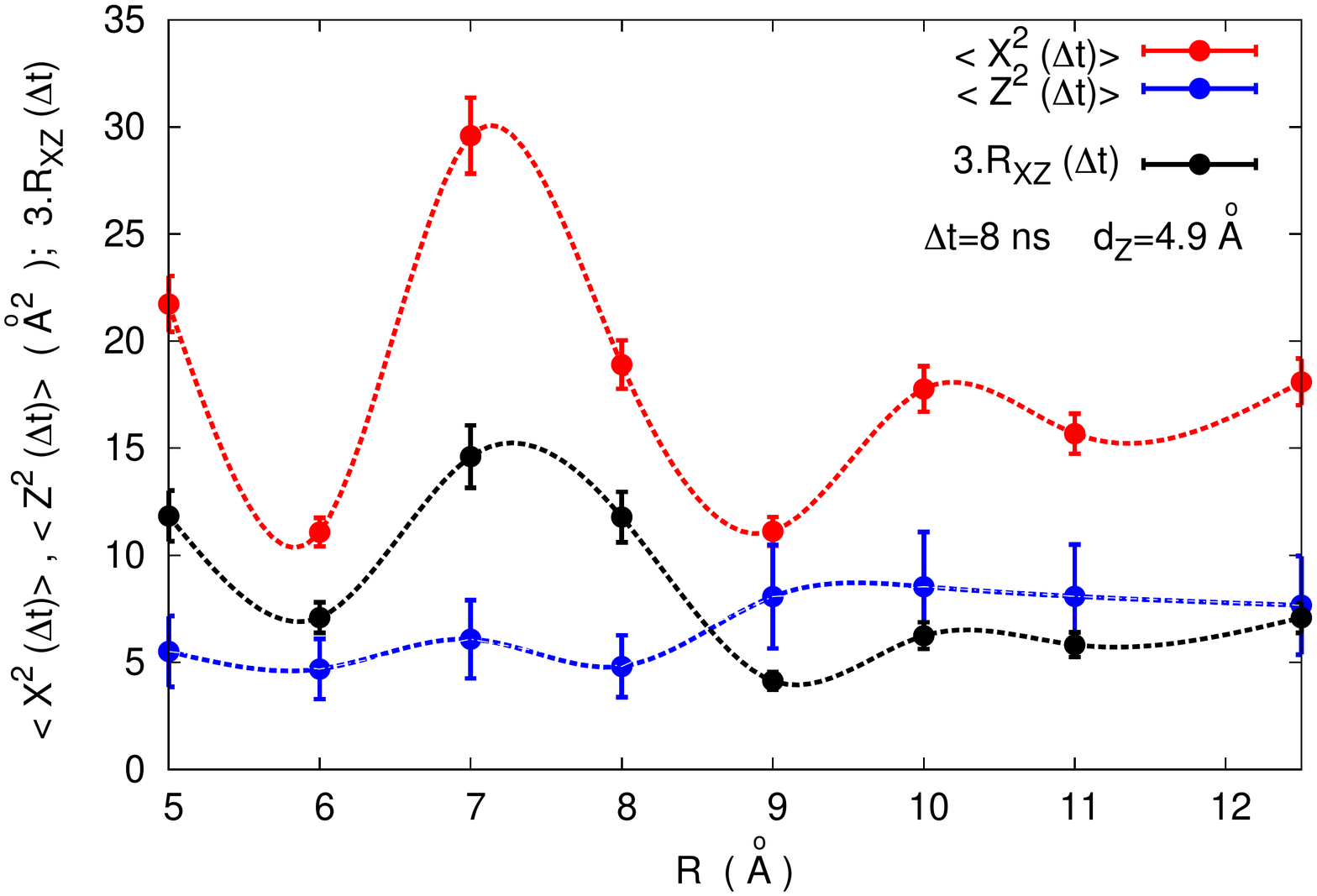}

\caption{(color online)  Ratio $R_{XZ}=<X^{2} (\Delta t)>/<Z^{2} (\Delta t)>$ of the motor's mean square displacement in the $X$ direction (i.e. in the direction of its main axis) and in the $Z$ direction (perpendicular to its plan) for a time lapses $\Delta t=8 ns$, together with the mean square displacements in the $X$ and $Z$ directions.
See Figure \ref{f0} for a representation of the motor and axes. 
}
\label{ff10}
\end{figure}

Now we will investigate how the direction of motion is modified inside  nano-pores.
For large time scales the motor has no choice but to move along the $z$ axis of the pore.
But does the layering and the pore walls also guide the motor at short time scales ?
To solve that question, in Figure \ref{ff10} we show the ratio $R_{XZ}=<X^{2} (\Delta t)>/<Z^{2} (\Delta t)>$ between its mean square displacement in its main axis direction $X$ and in the perpendicular $Z$ direction.
We do not take into account the $Y$ direction in this calculation, because the motor moves only slightly in that direction.
We also show in the same Figure the displacements $<X^{2} (\Delta t)>$ and $<Z^{2} (\Delta t)>$ for comparison.

When the motor is active Figure \ref{ff10} shows oscillations on the orientation of the motor's displacement versus the pore radius for short and large time scales.
  $R_{XZ}$ mostly oscillates between $2$ and $8$, while the orientation is almost constant for a passive motor with $R_{XZ}$ $ \approx 6$.
The orientation of the motor's displacement  therefore depends on the pore radius and activation, similarly to the diffusion coefficient.  
The Figure shows that the oscillations of the orientation $R_{XZ}=<X^{2}>/<Z^{2}>$ are mostly due to the oscillation of the displacement $<X^{2}>$ in the main motor's axis direction and only weakly on the oscillations of displacements $<Z^{2}>$ in the perpendicular direction.
For the two smallest activations studied in Figure \ref{ff10} ($d_{Z}=3.3$\AA\ and $3.8$\AA) we find two optimal pore radii $R=8$\AA\ and $11$\AA, that do not vary much for various activations.
For the largest activation ($d_{Z}=4.9$\AA) the optimal radii are shifted to $R=7$ and $10$ \AA, while for an activation in between ($d_{Z}=4.2$\AA), the shift is larger with optimal values are $R=6$ and $10$ \AA.

Figure \ref{ff10} also shows that the motor's displacement orientation is optimal for an activation $d_{Z}=3.8$ \AA\ and a pore radius of $8$\AA, while larger but less oriented displacement in the $X$ direction occur at larger activations. We notice that this activation was also found optimal in the bulk in a previous paper\cite{previous}.

\subsection{Average position of the motor within the pore compared with medium's layering}

\begin{figure}
\centering
\includegraphics[height=5.85 cm]{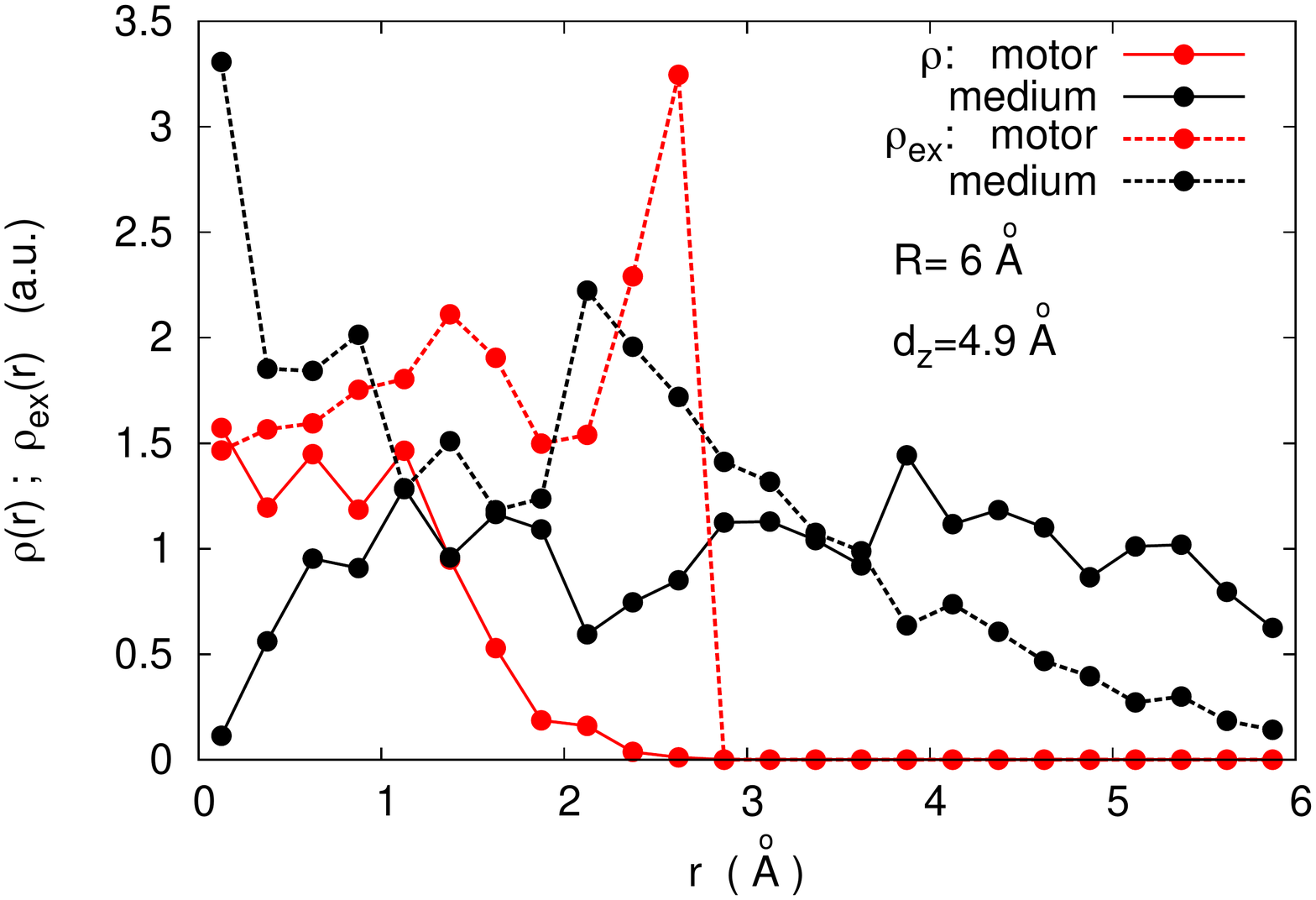}
\includegraphics[height=5.85 cm]{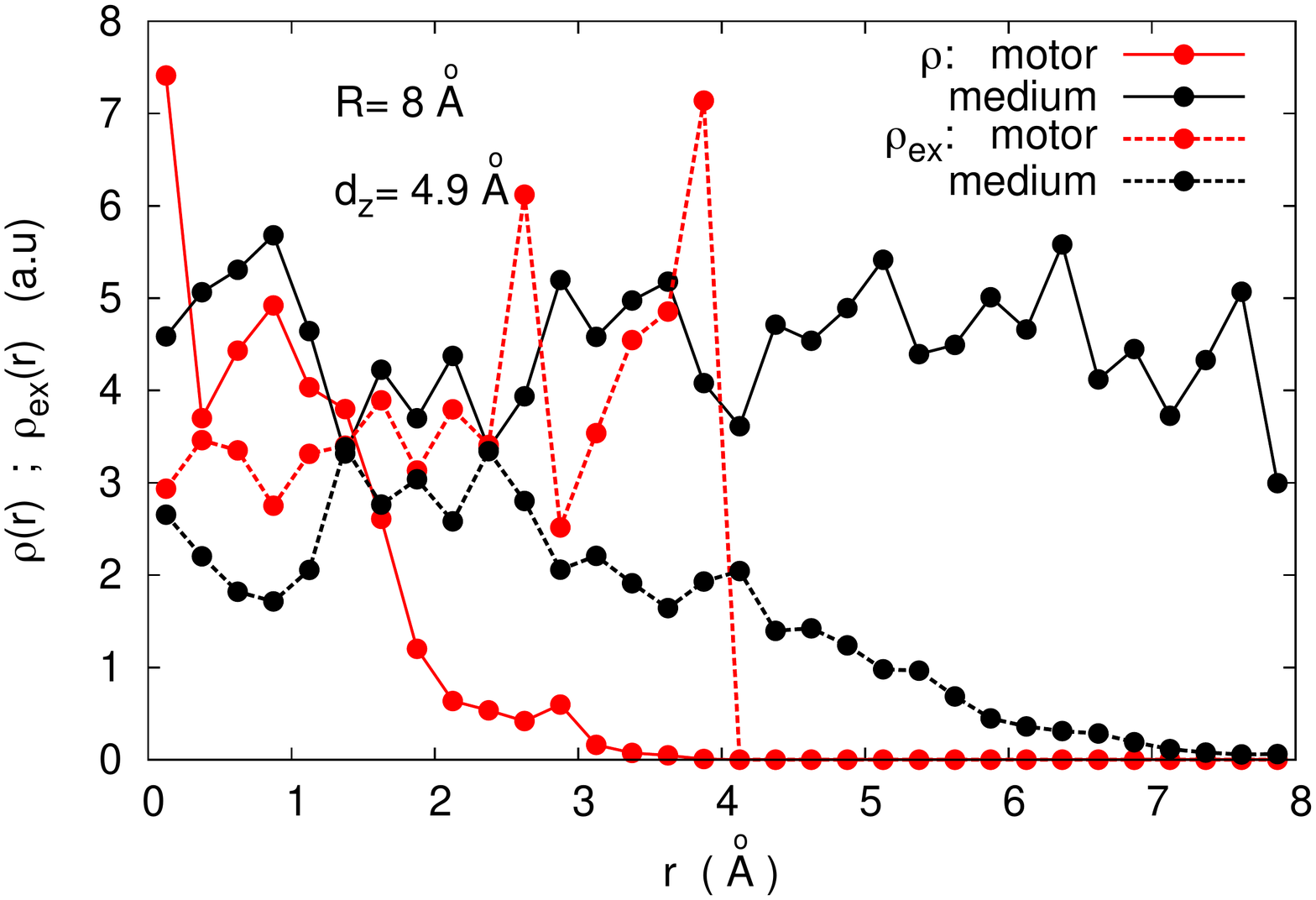}
\includegraphics[height=5.85 cm]{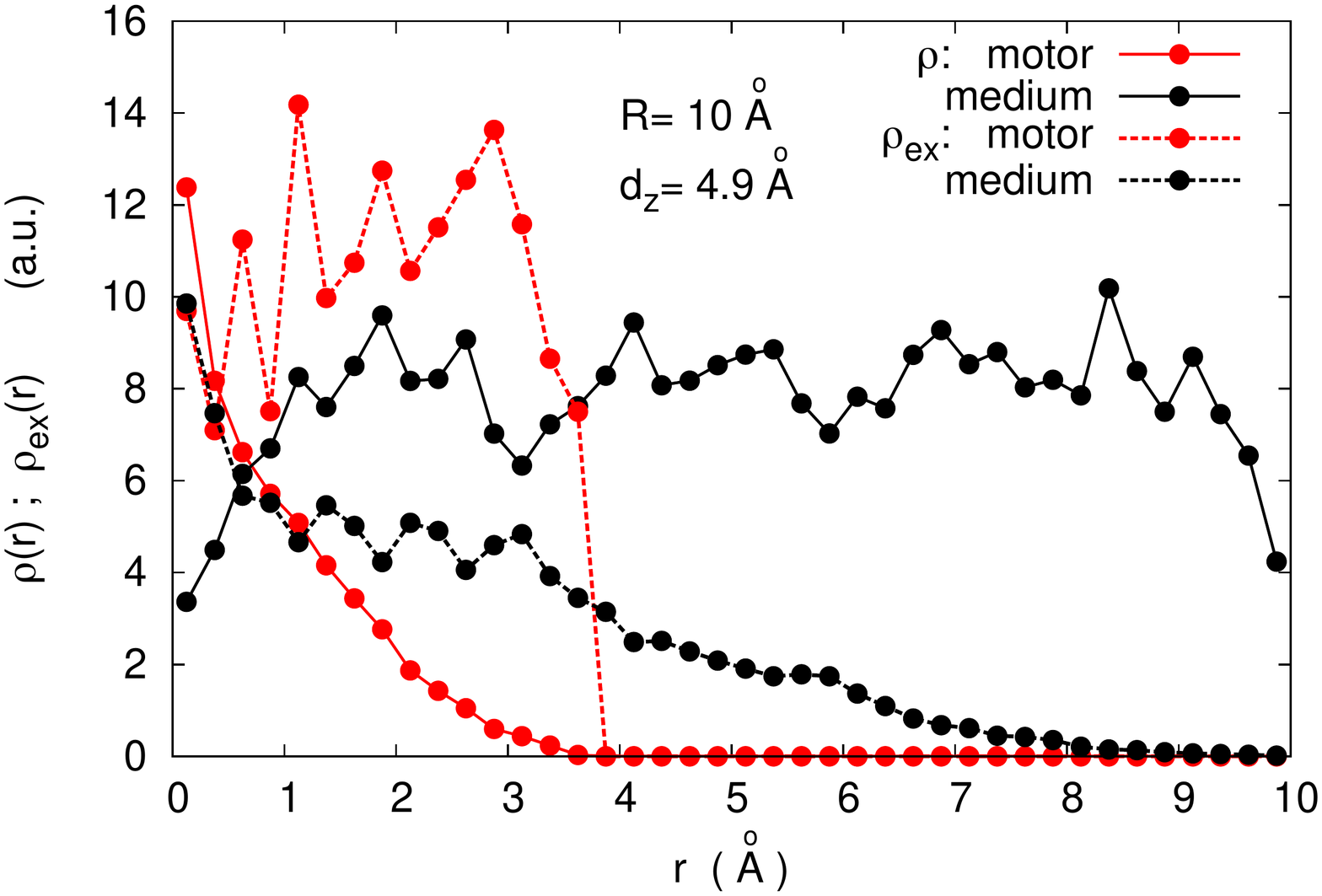}
\includegraphics[height=5.85 cm]{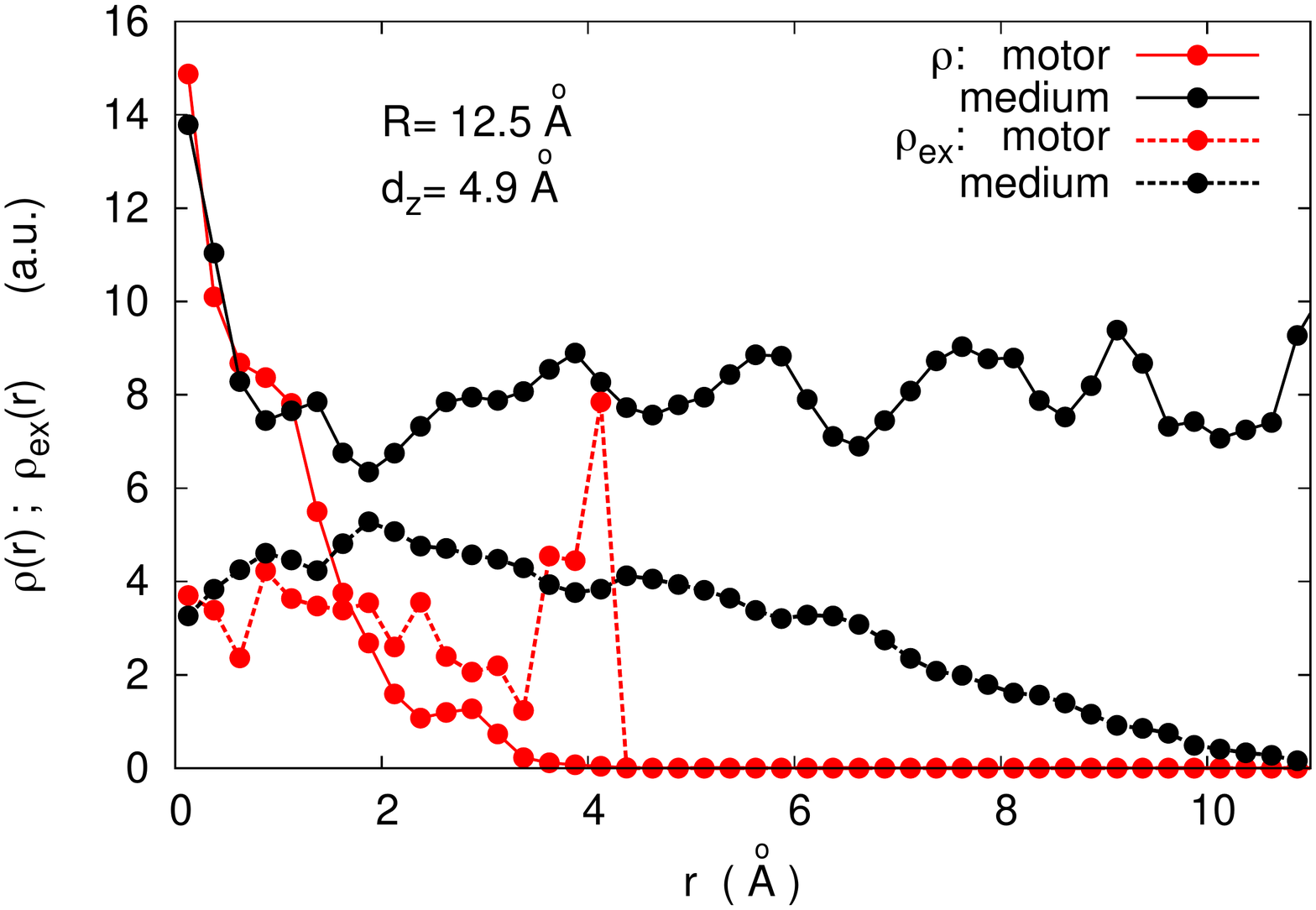}

\caption{(color online) Density distribution (continuous lines) of the motor (red) and medium (black) molecules,  together with the corresponding excitation (elementary diffusive motions) density distributions (dashed lines). 
The activation is constant $d_{Z}=4.9$ \AA\ while pore radii vary from $R=6$ \AA\ to $12.5$ \AA. 
}
\label{f5}
\end{figure}

\begin{figure}
\centering
\includegraphics[height=5.7 cm]{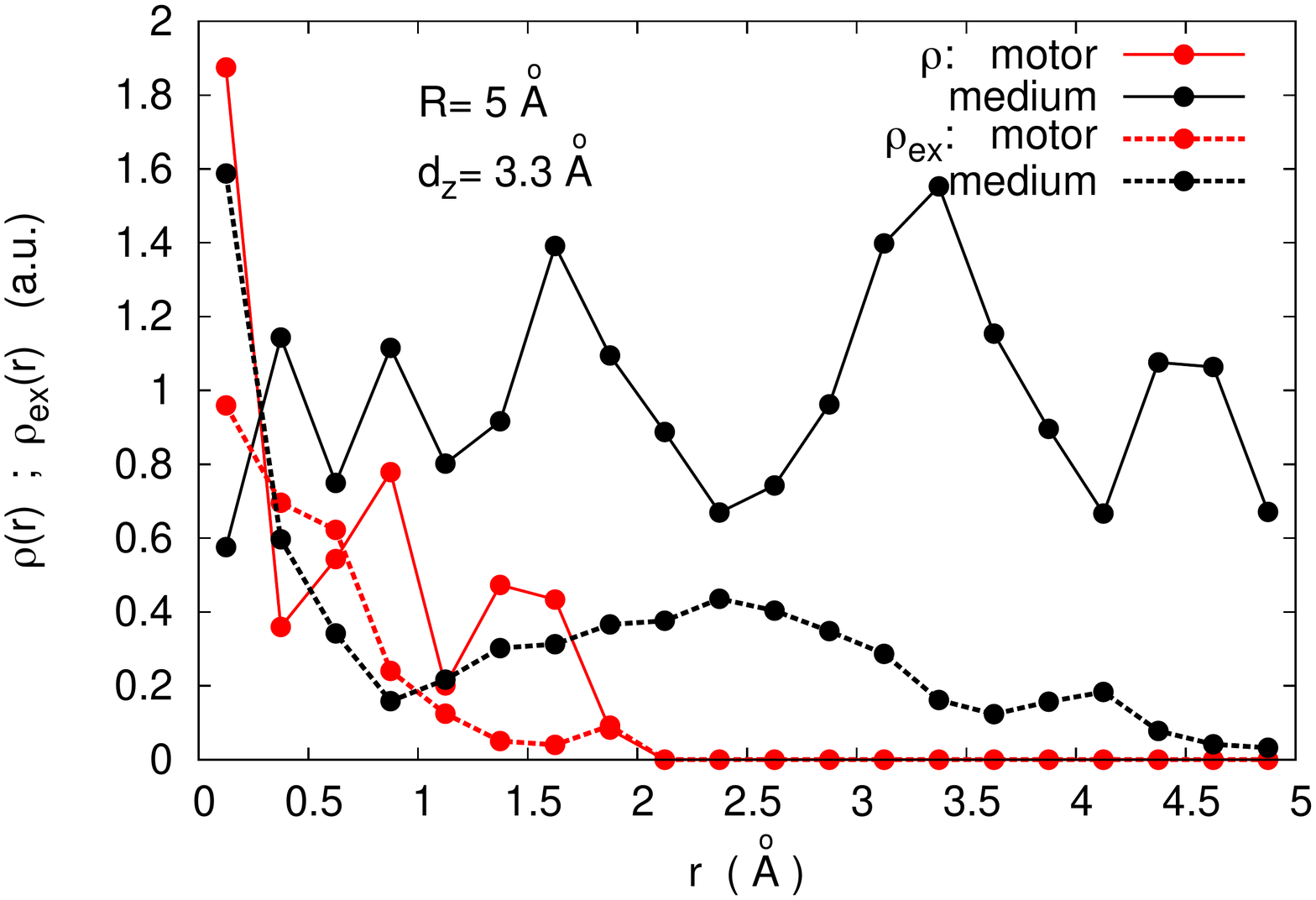}
\includegraphics[height=5.7 cm]{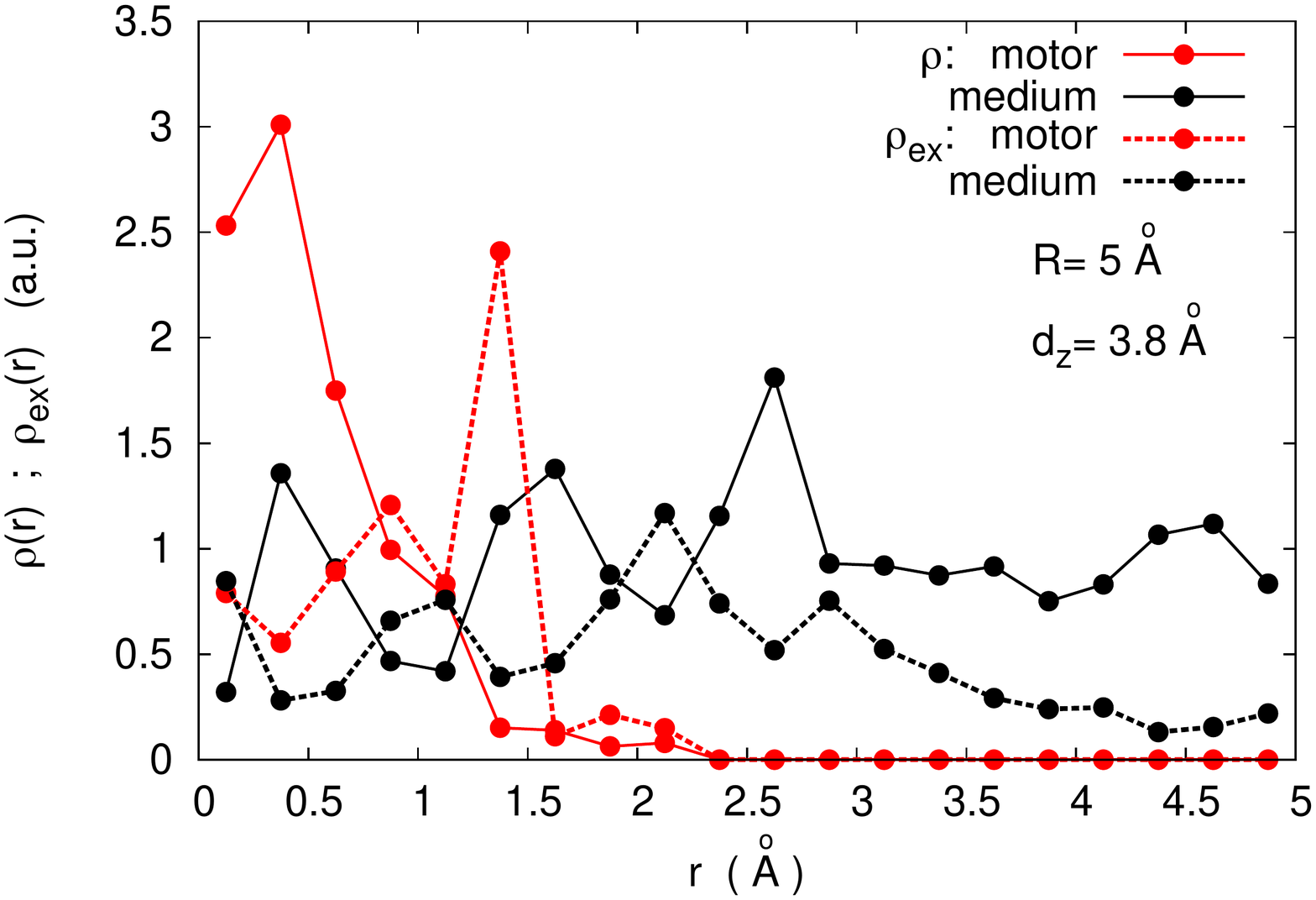}
\includegraphics[height=5.7 cm]{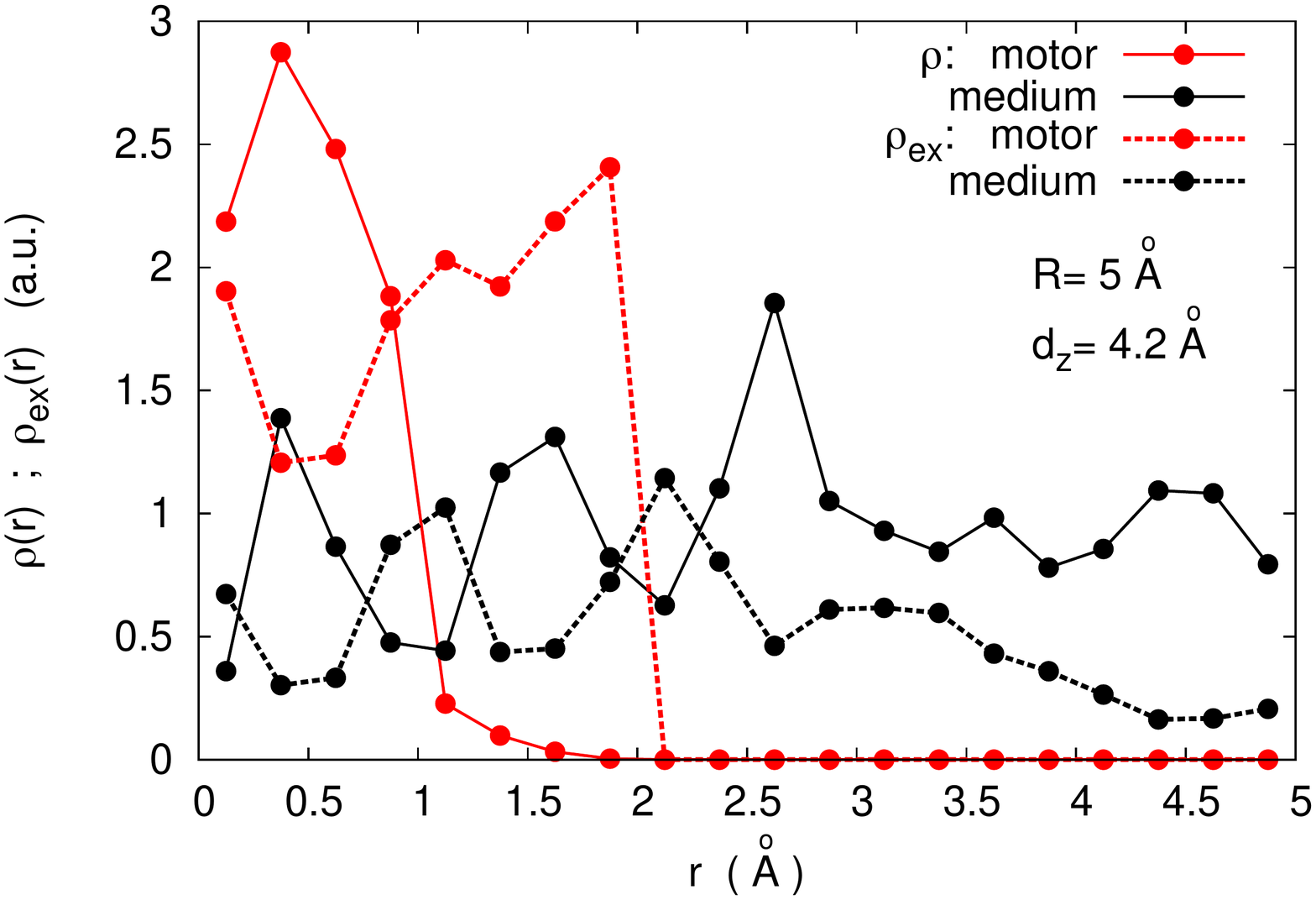}
\includegraphics[height=5.7 cm]{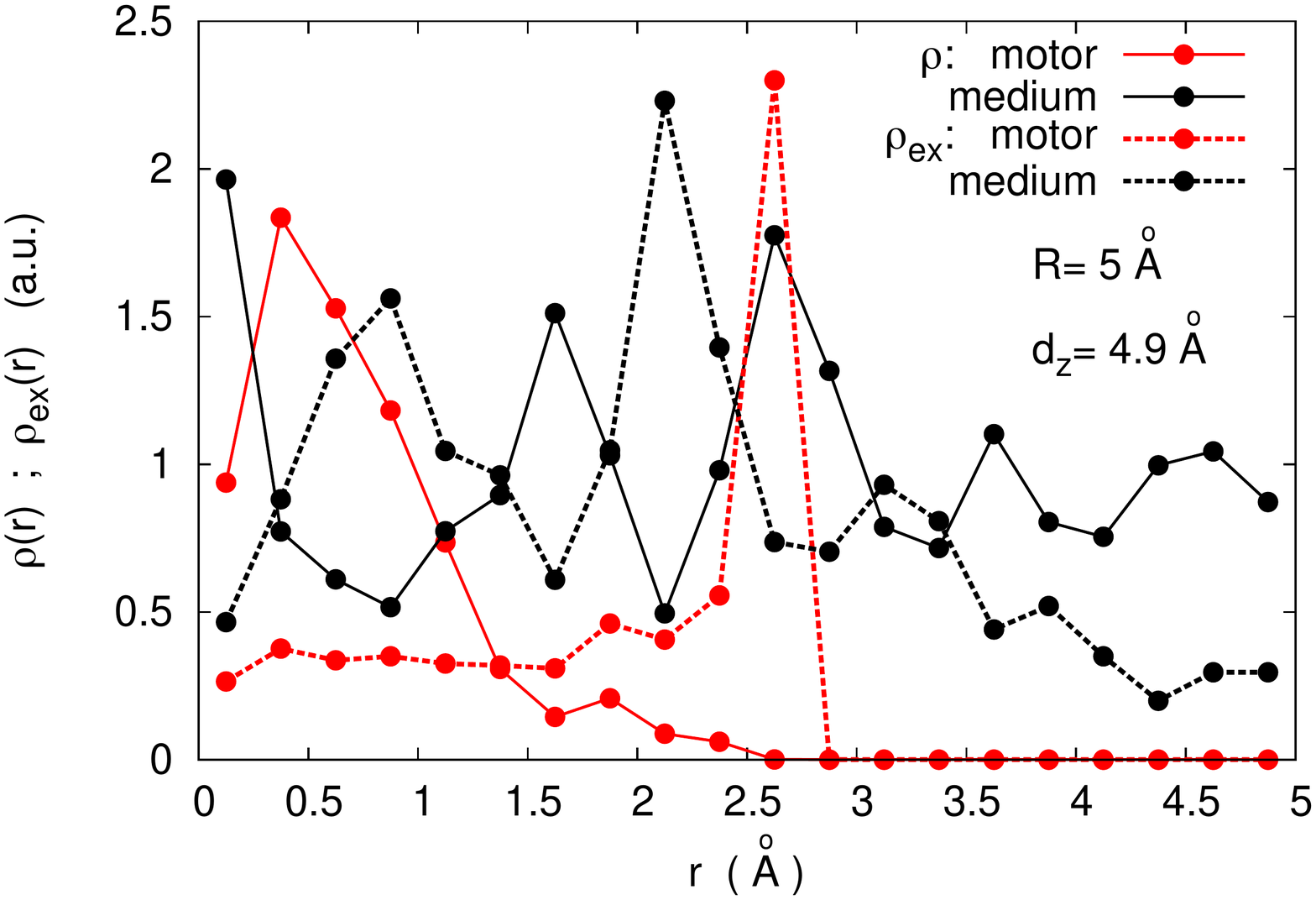}

\caption{(color online) Density distribution (continuous lines) of the motor (red) and medium (black) molecules,  together with the corresponding excitation (elementary diffusive motions) density distributions (dashed lines). 
The pore radius is constant $R=5$ \AA\ while the activation varies.
}
\label{f5b}
\end{figure}

\begin{figure}
\centering
\includegraphics[height=5.85 cm]{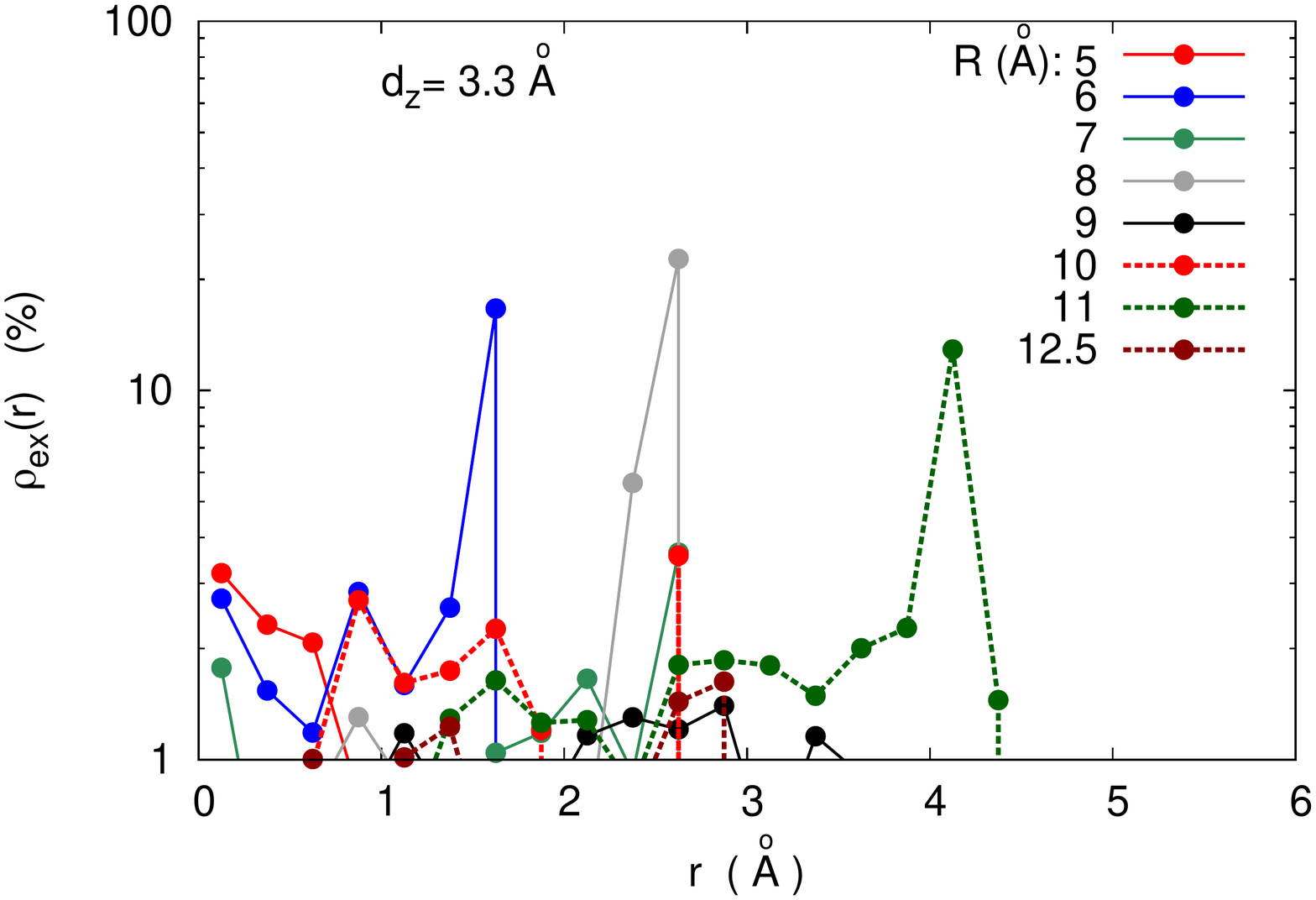}

\caption{(color online)  Motor's excitation (elementary diffusive motions) density distributions for various pore radii and an activation $d_{Z}=3.3$\AA.
Notice that the larger spikes are found for  optimal radii for motor's displacement and orientation: $R=6$, $8$ and $11$ \AA. 
}
\label{f5c}
\end{figure}

To better understand the behavior of the motor inside the pores, we compare in Figures \ref{f5} and \ref{f5b} the density distribution of the motor and medium's molecules together with the excitation densities, and the excitation densities for various radii in Figure \ref{f5c}. Excitations are here defined as molecules that move more than $1$ \AA\ in a $10 ps$ time lapse.
More precisely, in Figure \ref{f5} we focus on the radius effect while in Figure \ref{f5b} we show the activation effect for a constant pore radius.

We observe that for all pores studied the motor is mostly located at the pore center while medium molecules fill the whole pore, however with layering.
The Figures also show that the oscillations of the medium's excitations are in opposition with the medium's density oscillations for the whole set of results displayed in Figure \ref{f5}, showing that the medium's elementary displacements are larger when the local density is small.
We observe in Figure \ref{f5b} a modification of the layering, motor's location and excitation densities with the activation coefficient $d_{Z}$

Unexpectedly, the Figures show that the motor's center of mass location probability density tends rapidly to zero at a certain distance from the pore center.
The density location of the motor is restricted to a location ranging from $3$ to $4$ \AA\ from the pore center. 
This result suggests that due to the large size of the motor and its periodic folding, the pore confinement restricts its position more tightly than the position of smaller medium molecules. 
Now investigating the motor's excitation distribution, we  observe for most radii and activations, an excitation spike located at the very end of the motor's density distribution.
These results suggest that the motor bounces on the pore's walls, leading to the spike of excitation and decreasing the probability to be near the wall, in agreement with the previous remark.
Eventually, Figure \ref{f5c} shows that the excitations spikes are maxima for the pores radii where we found the maxima of displacements and orientation.
This result suggests that the bouncing of the motor leads to an important contribution to the motor's displacement.  
Following that picture, Figure \ref{f5} shows  proportionally larger spikes for small pores, in agreement with the observed increase of the activation effect for small pores.

\section{Conclusion}

In this work we used molecular dynamics simulations to investigate the displacement of a simple butterfly-like molecular motor inside nanopores of various radii.
We tested different activations of the medium inside the pore by varying the motor's opening angles. 
As previously found for the medium\cite{aa}, the activation decreases strongly the confinement's hindering of the motor's motion, in particular in small pores.
The optima of the motor's displacements oscillate with pore sizes and that the optimal radii  depend  on the activation of the medium.
Therefore it is possible to choose the activation that optimizes the motor's displacement for a given porous material. 
We interpret these oscillations as a consequence of  different layering of the medium inside different pores and for different activations, but we notice that bouncing motions also contribute differently for different pore radii.
Eventually we found that the motor was mainly located in the center of the pore for the whole set of pores tested.
In connection with that result we found  spikes in the density of elementary motions when the motor goes away from the pore center, revealing important bouncing motions on the pore walls.  
Work is in progress to better understand the appearance in this simple system of these bouncing motions that are reminiscent of much more complex biological motor proteins.

\vskip 0.5cm

{ \bf \large  Conflict of interest}

There are no conflict of interest to declare.
\vskip 0.5cm

{ \bf \large  Data availability}

The data that support the findings of this study are available from the corresponding author upon reasonable request.
\vskip 0.5cm


\end{document}